# Autocorrelated errors explain the apparent relationship between disapproval of the US Congress and prosocial language

Alexander Koplenig

Institute for the German language (IDS), Mannheim, Germany.

(1) claim that there is a linear relationship between the level of prosocial language and the level of public disapproval of US Congress. A re-analysis demonstrates that this relationship is the result of a misspecified model that does not account for first-order autocorrelated disturbances. A Stata script to reproduce all presented results is available as an appendix.

(1) claim that there is a linear relationship between the level of public disapproval of US Congress (disapproval) and the level of prosocial language within each month of Congress (prosocial-language). (1) fit a simple time-series regression that can be written as (2):

$$y_t = \beta_0 + \beta_1 x_{1_t} + \varepsilon_t$$

where $y_t$ represents the level of disapproval in $t$ and $x_{1t}$ is the level of prosocial-language, $\beta_0$ is the regression constant and $\beta_1$ is the regression coefficient, $\varepsilon_t$ is the error term. On that basis, (1) argue that there is a correlation between disapproval and prosocial-language ($r = 0.55$, $p < 0.001$). However, OLS analysis assumes that there is no autocorrelation between the residuals ($Cov(\varepsilon_s, \varepsilon_t) = 0$ for all $s \neq t$). In this context, first-order autocorrelation $\varepsilon_t$ can be written as:

$$\varepsilon_t = \rho \varepsilon_{t-1} + \eta_t$$

where $\eta_t$ is a white-noise process. In the presence of first-order autocorrelation, the OLS estimators are biased and lead to incorrect statistical inferences (3).

Using the data made available by (1), Fig. 1 plots the residuals of a regression of disapproval on prosocial-language against the lagged residuals[1]. A visual inspection of the plot implies that there is strong first-order autocorrelation. The alternative Durbin-Watson statistic supports this impression ($d(12) = 473.98$, $p < 0.001$).

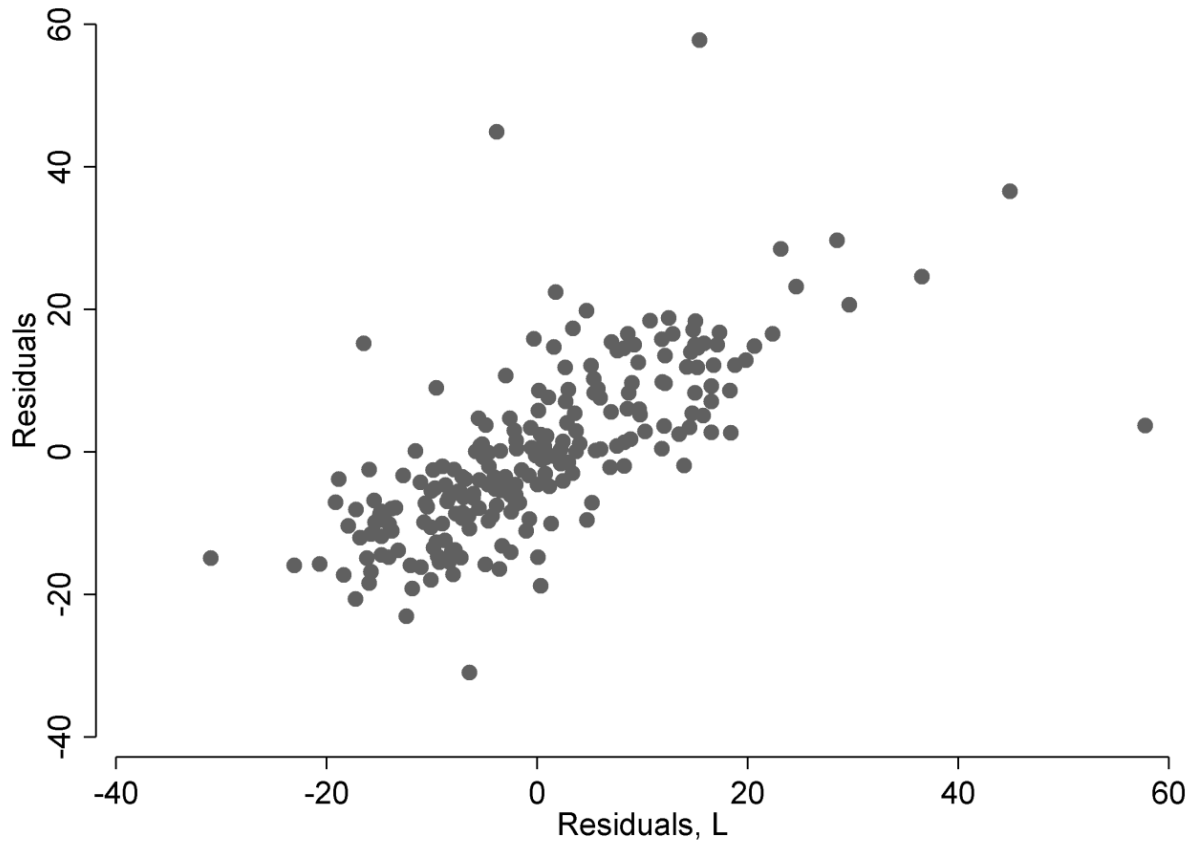

**Fig. 1: Current residuals against lagged residuals of an OLS regression of the level of disapproval on the level of prosocial-language.**

An augmented Dickey-Fuller test (2) implies that both series are non-stationary ($p = 0.87$ for disapproval and $p = 0.34$ for prosocial-language; both with 12 lags for monthly data). Regressing one non-stationary series on another non-stationary series leads to a spurious model (3). To obtain (weakly) stationary series, one can take first differences of the two series and compare month-to-month changes instead of the levels, using the following notation:

$$\Delta y_t = \beta_0 + \beta_1 \Delta x_{1_t} + \varepsilon_t$$

---

[1] For this analysis, missing values due to small samples sizes (1) in the series of the level of prosocial language where linearly interpolated.

The first difference of both series are (weakly) stationary (approximate $p$s < 0.001). Regressing monthly changes of disapproval on monthly changes of prosocial-language leads to a correlation between both series that is virtually zero ($r = -0.07$, $p = 0.27$). The alternative Durbin-Watson statistic is now much smaller ($d = 33.93$), but still significant at $p < 0.001$. Therefore, we can estimate a first-order ARMAX model, where the regression errors can be written as (4):

$$\varepsilon_t = \rho \varepsilon_{t-1} + \theta \eta_{t-1} + \eta_t$$

where $\rho$ is the autoregressive parameter, $\theta$ is the moving average parameter and $\eta_t$ is a white-noise process . Such a model with robust standard errors yields an insignificant negative effect ($p = 0.42$) of the first difference of disapproval on the first difference of prosocial-language. A joint test of the significance of the ARMA parameters shows that both parameters are not significantly different from zero which indicates that the chosen ARMA specification is correct ($\chi^2 = 67.25$, $p < 0.001$).

This re-analysis casts doubt on the results of (1).

Appendix

```
/* Stata do file for:
Autocorrelated disturbances explain the apparent relationship
between disapproval of the US Congress
last checked: 06/02 /2015

download data here:
https://osf.io/94gc5/?action=download&version=1
*/

import excel "F:\Public Data.xlsx", sheet("Summary Variables") cellrange(A4:T234) clear

drop if A==.
/* generate date variable */

gen mdate=ym(A,B)

/* prosocial words */
gen double prosocial=H

/* congress approval */
gen congress=O

keep mdate pro* congress
order mdate

tsset mdate, m

/* test if correlation are equal to Frimer et al. */

pwcorr congress prosocial*, sig

/* interpolate for missing values */

ipolate prosocial mdate, gen(iprosocial)

/* OLS regression */
reg congress iprosocial

predict residuals, residuals

/* Fig.1 */

scatter residuals L.residuals, ///
```

```
        scheme(s2mono) graphregion(color(white)) ///
        yscale(nofextend) xscale(nofextend) ylabel(, nogrid)
        graph export 1.tif, height(2000) replace
window manage close graph

estat durbinalt, lags(12)

/* test for a unit root */

dfuller congress, l(12)
dfuller iprosocial , l(12)

/* differencing */

dfuller D.congress, l(12)
dfuller D.iprosocial , l(12)

capture drop residuals

/* OLS regression */
reg D.congress D.iprosocial
predict residuals, residuals

estat durbinalt, lags(12)

pwcorr D.congress D.iprosocial, sig

/* ARIMA model */

capture drop residuals
regress D.congress D.iprosocial
predict residuals, residuals
ac residuals, lags(20) note("")   name(ac, replace)  nodraw
pac residuals, lags(20) note("")   name(pac, replace) nodraw
graph combine ac pac

arima congress iprosocial, arima(1,1,1) vce(robust)

predict earma, residuals
pac earma
ac earma

/* joint test for significance */
```

test [ARMA]

exit

contact: koplenig@ids-mannheim.de